\documentstyle[emulateapj]{article}

\lefthead{Escala}
\righthead{Towards a 
Comprehensive Fueling-Controlled Theory on the Growth of Massive Black Holes and Host Spheroids.}

\def\msun{M_{\odot}}

\begin{document}

\title{Towards a 
Comprehensive Fueling-Controlled Theory on the Growth of Massive Black Holes and Host Spheroids.} 
\author{Andr\'es Escala}
\affil{Kavli Institute for Particle Astrophysics and Cosmology}
\affil{Stanford University Physics Department / SLAC, 2575 Sand Hill Rd. MS 29, Menlo Park, CA 94025, USA.}

\begin{abstract}

We study the relation between nuclear massive black holes and their host spheroid  gravitational potential. Using AMR numerical simulations, we analyze how gas is  transported in the nuclear (central kpc) regions of galaxies. We study the  gas fueling onto the inner accretion disk (sub-pc scale) and the star formation in a massive nuclear disk like those  generally found in  proto-spheroids (ULIRGs, SCUBA Galaxies). These sub-pc resolution simulation of gas  fueling that is mainly depleted by star formation naturally satisfy  the `$\rm M_{BH}$ - $\rm M_{virial}$'   relation, with a  scatter considerably less than the observed one. 
We found a generalized version of  Kennicutt-Schmidt Law for starbursts is satisfied, in which the total gas depletion rate ($\rm \dot{M}_{gas} \, = \, \dot{M}_{BH} \, + \,  \dot{M}_{SF}$) is the one that  scales as $\rm M_{gas}/t_{orbital}$. We also found that the `$\rm M_{BH}$ - $\sigma$' relation is a byproduct of the `$\rm M_{BH}$ - $\rm M_{virial}$'  relation.

\end{abstract}


\section{Introduction}

In the past years, it has been found that most nearby massive spheroids (elliptical and spiral bulges) host nuclear massive black holes (MBH) (Kormendy \& Richstone 1995), whose  masses correlates with the host spheroid properties. Two correlations arises as the more relevant  links between MBH and their hosts.  The first one is that  the masses of black holes correlates with  the mass of the host (`$\rm M_{bh}-M_{bulge}$' relation; Marconi \& Hunt 2003; Haring \& Rix 2004; Lauer et al. 2006), and in the second one the masses correlates with average random velocities of the stars in their host (`$\rm M_{bh}-\sigma$' relation; Ferrarrese \& Merritt 2000; Gebhardt et al. 2000). Examples of the other relations can be found in Graham \& Driver (2007) and references therein. Several theories have been proposed  to clarify the origin of these relations  (Silk \& Rees 1998; Blandford 1999; Fabian 1999; Burkert \& Silk 2001; Zhao, Haehnelt \& Rees 2002; King 2003; Adams et al. 2003; Miralda-Escaud\'e \& Kollmeier 2005; Sazonov et al. 2005; Begelman \& Nath 2005; Escala 2006, Peng 2007). However, very little numerical work  has been devoted to test  their key hypothesis and assumptions in order  validate these theories.


The study of MBH growth by gas accretion is usually  focused  on the study of accretion disks. However, these accretion disks are Keplerian by nature and therefore  have neglible masses compared to that of the   MBH. They must be continously replenished, otherwise the mass  of the MBH  will not  have a considerable growth. The key question in the growth of MBHs by accretion, is  how to remove the large angular momentum of gas in a galaxy in order to funnel it into the accretion disk in the central sub-pc region: the so called  `Fueling Problem'. This fueling of gas that is an unavoidable step in the growth of MBHs by accretion, is by itself   a galactic problem. Therefore, is  a natural candidate for being the responsible for the correlation between the MBH masses and the galactic properties of their hosts, without requiring   any extra ad-hoc hyphotesis such as huge outflows, fine tunning on the gas mass to stellar mass ratio, etc.

There are several mechanisms for  fueling gas down to  kpc scales such as  galaxy mergers/interactions, bars and resonances (see Shlosman, Begelman \& Frank 1990; Wada 2004 for a review). Gravitational torques in galaxy mergers arise as  the dominant process for fueling large amounts of gas down to the central few hundred parsecs and for triggering most of the MBH growth.  In a merger, after a violently relaxed core is formed at the center, most of the gas  will settle  in a nuclear disk (with typically several hundred parsecs in diameter) 
  that is  rotationally supported against the overall (gas + stars) gravitational potential (Barnes 2002; Kazantzidis et al. 2005; Mayer et al. 2007). The MBHs will migrate to the center, and merge in a timescale relatively short compared to the lifetime of the nuclear disk (Escala et al. 2004, 2005; Dotti et al. 2006).

In this paper, this massive nuclear disk with a central MBH is the starting point of our study. We focus on  the mass fueling onto  the inner accretion disk around a MBH (from the few hundred pc  down to sub-pc scale), with special interest  on the role of the spheriod gravitational potential in the mass transport. For that reason,  we have designed a  simple and clear experiment to  study  of the mass transport in a spheroid-dominated system. Here we analyze the mass transport in the same exact disk in rotational support with different background spheroid, ranging from simulations where the spheriod  represent a typical bulge to  ones that represent  a big elliptical. With this experiment we can get insight of  the fate of a fix amount of gas in a given galaxy in the universe
, how much of that gas will form stars and how much will be feeded to the central MBH depending on the spheroid that host it. 


Using simple analytical models Escala (2006), hereafter E06, analyzed how gas is expected to be transported in the nuclear regions of galaxies. For disks where  the expected gas lifetime  is given by the  Kennicutt-Schmidt Law, this simple models naturally led E06 to the `$\rm M_{BH}$ - $\rm M_{virial}$'  and `$\rm M_{BH}$ - $\sigma$' relations. One of the main motivations  of this paper is to test the conclusions of E06 based on simple models by considerably  more realistic adaptive mesh refinement (AMR) simulations.

We start with  a description of the assumed initial conditions, the model setup and the numerical results in \S 2. We continue with the analysis of the origin of the `$\rm M_{BH}$ - $\rm M_{virial}$'   relation in \S 3. In \S 4 we compare the `$\rm M_{BH}$ - $\rm M_{virial}$'  and `$\rm M_{BH}$ - $\sigma$' relations. In \S 5 we discuss the implications of our work. Finally, our summary is presented in \S 6.

\section{Numerical Setup}

The model consist in a massive  gaseous disk  around a MBH  that is embedded in a stellar spheroid. The spheroid is modeled by a  time-independent external potential. We use the gravitational potential of an isothermal sphere for $\rm r \geq 100 pc$, and the potential of an homogenous sphere for $\rm r < 100 pc$. We solved the hydrodynamic and Poisson equations  using the ENZO adaptative mesh refinement hydrodynamics code (Bryan and Norman 1997). The adaptative cartesian grid covers a  $\rm 1 kpc^{3}$  region around the galactic center, with a  spatial resolution of 0.24 pc. Radiative gas losses  from an optically thin plasma follows the cooling curve of Sarazin \& White (1987) down to temperatures of $\rm 10^4$K. We use outflow (zero gradient) boundary conditions and  in our simulations we use the following units: [Mass] = $\rm 1.76 \times 10^{10}\msun$, [Velocity] = $\rm 276 km s^{-1}$,  and  [Distance] =  1 kpc.


For modelling star formation, the following criteria were used to decide whether a grid cell would produce a star (Cen \& Ostriker 1992; O'Shea et al. 2004): (1) the gas density in that grid cell exceeds a threshold density, (2) the mass of gas in the cell exceeds the local Jeans mass, (3) there is convergent flow (i.e., $ \nabla \cdot v < 0$), and (4) the cooling time is less than the dynamical ($\rm t_{cool} < t_{dyn}$), or the gas temperature is at the minimum allowed value. If a grid cell meets all the previous criteria, then some gas is converted into a `star particle'. The mass of this star particle is calculated as
\begin{equation}
\rm m_{\star} = \epsilon \frac{\Delta t}{t_{dyn}} \rho_{gas} \Delta x^{3} \, ,
\end{equation}
where $\rm \epsilon$; is the star formation efficiency  per dynamical time (equals to 0.1), $\rm \Delta t$ is the size of the time step, $\rm t_{dyn}$ is the time for dynamical collapse, and $\rm \rho_{gas}$ is the gas density. This set of conditions has one extra criterion added to it: even if a cell fulfills all of the previous four criteria, a star particle will not be formed if its mass is less than a minimum star particle mass $\rm m_{\star}^{min}$. In our simulations, the value for $\rm m_{\star}^{min}$ used was $\rm 10^{3} M_{\odot}$. In the case in which this criteria is the only mechanism preventing a star particle from forming, a bypass exists that allows a star particle with mass less than $\rm m_{\star}^{min}$ to form if $\rm m_{\star}^{min}$ is greater than  80\% of the mass in the cell. To model the star formation in a molecular cloud, which will typically spread out over a dynamical time, the star particle's mass  is spread out over a time $\rm \tau = max(t_{dyn}, 1 Myr)$.

The massive black hole with its accretion disk is modelled as a sink particle that is originally placed at the center of the system. This sink particle accretes the gas that is  within 5 cells ($\rm \sim 1 pc$) around the center, and that has exceeded the  critical number density  $\rm n_{crit} = 7 \times 10^{7}  \, cm^{-3}$. 

The initial condition is a rotationally supported disk with a 1/r density profile,  600 pc of diameter and thickness of 30 pc. The total mass of the disk is $\rm M_{gas} = 3 \times 10^{8} \msun$ and the initial mass of the central black hole is $\rm M_{BH} = 10^{6} \msun$. The initial temperature is set to $\rm 10^{4}$K over the whole region. Random density and temperature fluctuations of less than 1\% are added to the initial uniform disk. We performed 6 different runs where we vary the mass of the bulge, in  such a way that the velocity dispersion of the bulge  $\rm \sigma$ is: 110, 163, 216, 270, 321 and 357 $\rm km \, s^{-1}$. For a bulge modeled by an isothermal sphere, our rotationally supported disks have flat rotation curves. For example  the run with the highest gas fraction $\rm M_{gas}/M_{star}$  has a rotation velocity $\rm  v_{rot}$ of  129 $\rm km \ s^{-1}$. This corresponds to the $\rm \sigma =  110 km \, s^{-1}$ run and the $\rm M_{gas}/M_{star}$ ratio is approximately 1/3 and  similar to those found in the most gas rich ULIRGs (Downes \& Solomon 1998). Table 1 lists the parameters used in these six  runs (note that $\rm M_{star}$ is the total stellar mass within the initial disk equivalent to the  applied external gravitational potential).

\begin {center}
\centerline{Table 1: Run Parameters}
\begin{tabular}{ccccc} \hline

RUN  &  $\rm \sigma[km \, s^{-1}]$ &  $\rm  v_{rot}[km \, s^{-1}] $  &  $\rm M_{gas}[\msun]$  &   $\rm M_{star}[\msun]$ \\ \hline \hline 1  &  110  & 129  & $\rm 3 \times 10^{8}$  &  $\rm  8.6 \times 10^{8} $    
\\  1.5  &  163 & 176  &  $ \rm 3 \times 10^{8}$ &  $\rm   1.9 \times 10^{9} $ \\  2  & 216 &  226  & $ \rm 3 \times 10^{8}$  &  $\rm 3.2 \times 10^{9} $  \\  2.5  & 270 &  278  & $ \rm 3 \times 10^{8}$  &   $\rm 5.1 \times 10^{9}$ \\  3  & 321 &  328  & $ \rm 3 \times 10^{8}$  &   $\rm 7.2 \times 10^{9}$ \\  3.5  & 357 &  364  & $ \rm 3 \times 10^{8}$  &   $\rm  8.9 \times  10^{9}$ 

\\ \hline

\end{tabular}
\end{center}

\subsection{Results}

\begin{figure}
\plotone{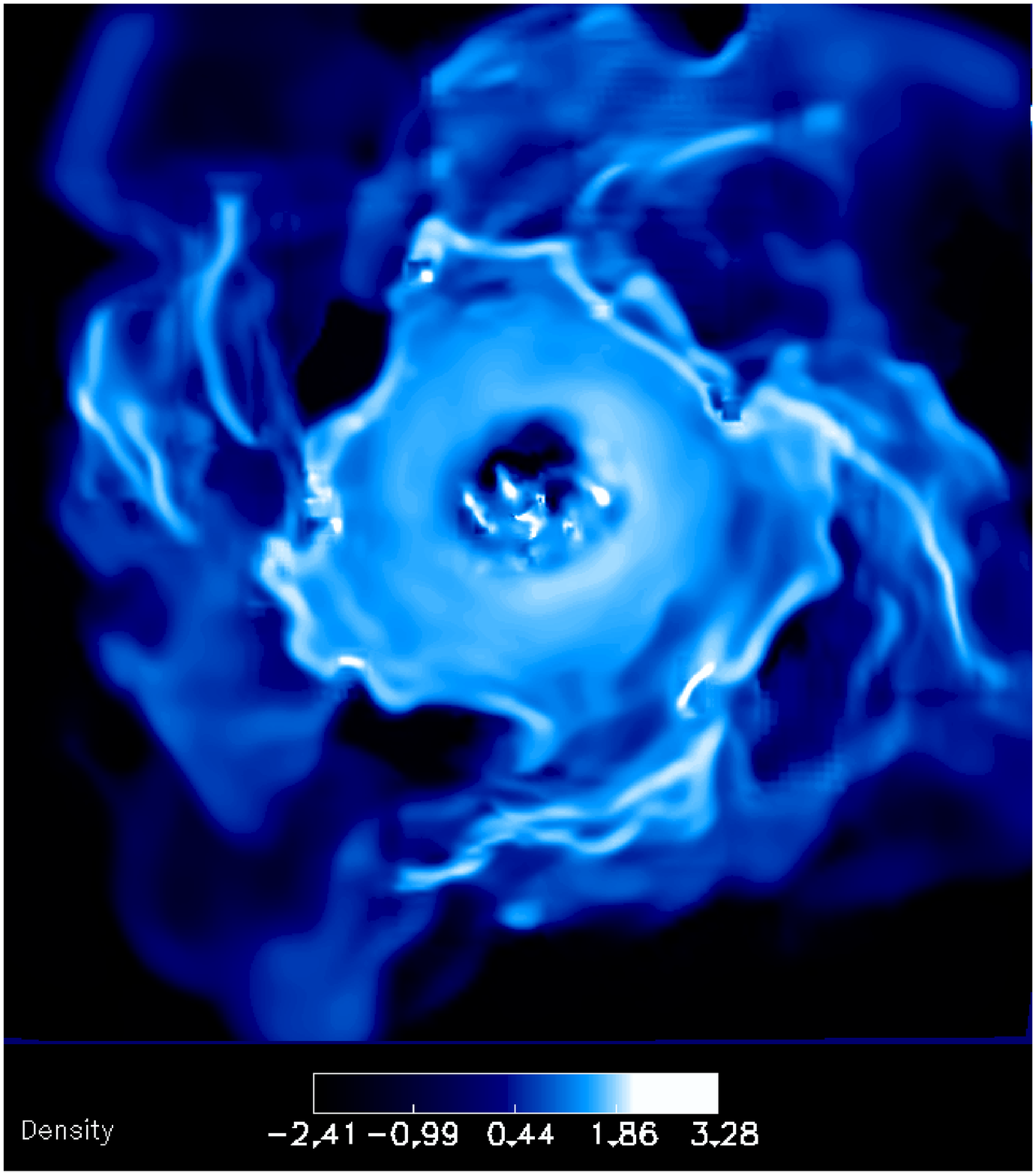}
\caption{Density distribution in the plane of the gas disk, in code units and coded on a logarithmic scale, at time t = 2.2Myr  for  the run 3.5. The plotted region in the x-y plane is 1kpc$\times$1kpc.
 The figure shows a complicated multiphase   structure,  characterized by high density clumps and filaments that  are embedded in a less dense medium.}
\label{f1}
\end{figure}

Figure $\ref{f1}$ illustrates a representative stage in the evolution of the system, showing the face-on density distribution at the plane of the disk (z=0) for  run 3.5, at the time t = 2.2 Myr. The plotted region in the x-y plane is 1kpc$\times$1kpc. The figure shows a complicated multiphase  
 structure,   in qualitative  agreement with  the findings of  Wada \& Norman (2002) and E06. The medium is characterized by high density clumps and filaments,  embedded in a less dense medium.

\begin{figure}
\plotone{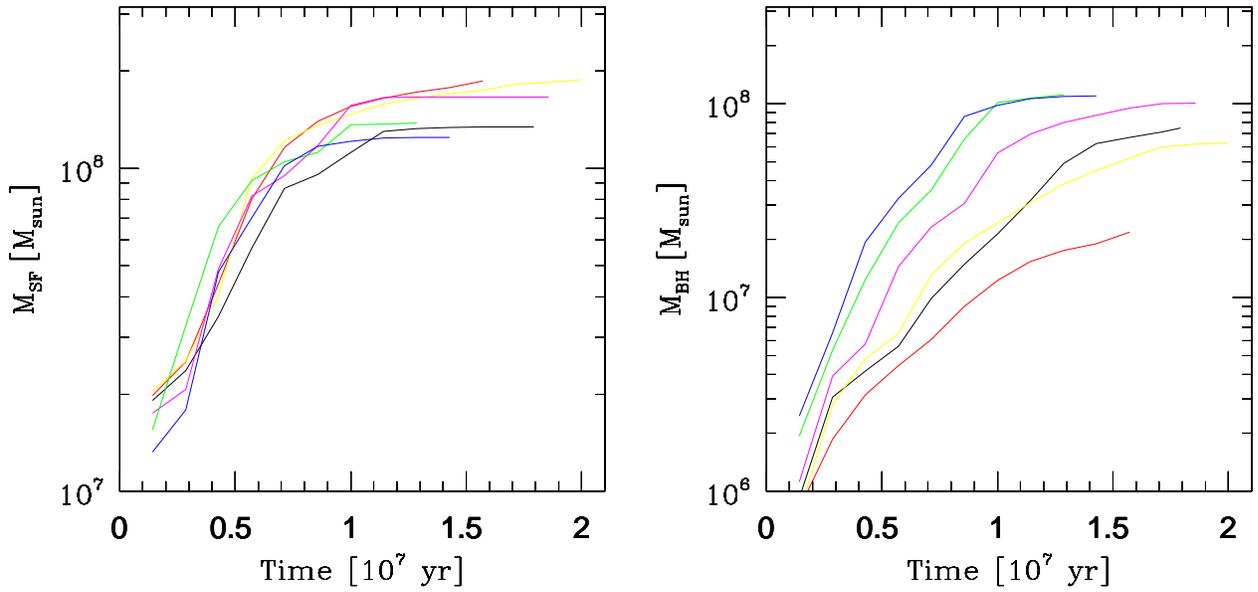}
\caption{(a) Time  evolution  of the total mass of stars formed, for the six different runs: 1 (red), 1.5 (yellow), 2 (black), 2.5 (magenta), 3 (green) and 3.5 (blue). (b) Time evolution of the total mass accreted by the MBH, $\rm \dot{M}_{BH}$. The colors in (b)  match those in (a).}
\label{f2}
\end{figure}

Figure $\ref{f2}$a shows the time  evolution  of the total mass of stars formed, for the six different runs: 1 (red), 1.5 (yellow), 2 (black), 2.5 (magenta), 3 (green) and 3.5 (blue).  In all the runs, the total mass of stars formed shows an approximately  exponential growth until reaches  saturation.  The figure doesn't shows a strong dependence on the background gravitational potential, which is the only parameter varied among the runs.

In Fig. $\ref{f2}$b, we plot the time evolution of the total accreted mass onto the MBH,  for the different runs and using the same color code as in  Fig. $\ref{f2}$a. This figure shows a strong dependence on the background gravitational potential, contrarily to that seen in the formation of stars.  The final MBH mass varies by an order of magnitude among the runs. In all the runs, the  total accreted mass  shows an approximately  exponential growth until reaches  saturation. 

\begin{figure}
\plotone{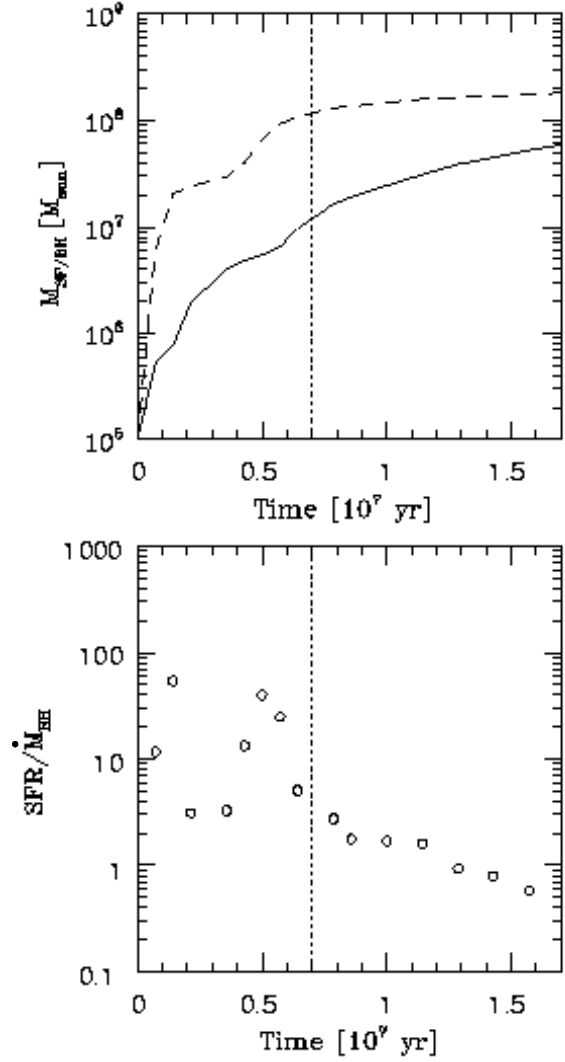} 
\caption{(a) Time evolution of  the total mass of stars formed (dashed line) and of the  mass accreted onto the MBH (solid line), for  the run 3.5. (b) The open circles shows the time evolution of the ratio between the star formation rate and the MBH mass accretion rate, which clearly decays after t = 7 Myr (marked with the dotted line).}
\label{f3}
\end{figure}

Figure  $\ref{f3}$a shows  a comparison of  the total accreted mass onto the MBH (solid line)  and the  total mass of stars formed (dashed line) as a function of time  for run 1.5. In the early evolution of the system, up to t = 7 Myr, we see  that the BH growth and the star formation in the host  co-evolves (i.e. $\rm \dot{M}_{BH} \propto \dot{M}_{SFR}$) as has been suggested by recent observations of proto-spheroids (Kauffmann \& Heckman 2005; Hao et al. 2007). This co-evolution arises naturally in this scenario where is the same massive disk the one responsible for the mass accretion rate onto the BH ($\rm \dot{M}_{BH}$) and for the star formation rate in the galaxy ($\rm \dot{M}_{SFR}$). Contrarily in the standard scenario where feedback  suppresses further star formation in the host galaxy, you expect non-correlation or even anti-correlation between $\rm \dot{M}_{BH} \,\, and \,\, \dot{M}_{SFR}$.


At later times in Fig.  $\ref{f3}$a, after  t = 7 Myr, the MBH mass tend to converge towards to  total mass of stars formed. This  change in   trend  is because the star formation tends to became inefficient  after that time as seen in  Fig. $\ref{f3}$b. This figure plots the time evolution of the ratio between the star formation rate and the MBH mass accretion rate, which clearly decays after t = 7 Myr. The origin of the decay is that as most of the gas gets consumed and the disk  become less dense and gravitationally more stable, therefore the gas avoids to forms stars and mainly  ends-up in the central BH. However, we don't think that this steep decay happens  in real starbursts because other processes not included in our simulations, like feedback from star formation or AGN, will continue producing inhomogeneities and triggering further star formation.

\begin{figure}
\plotone{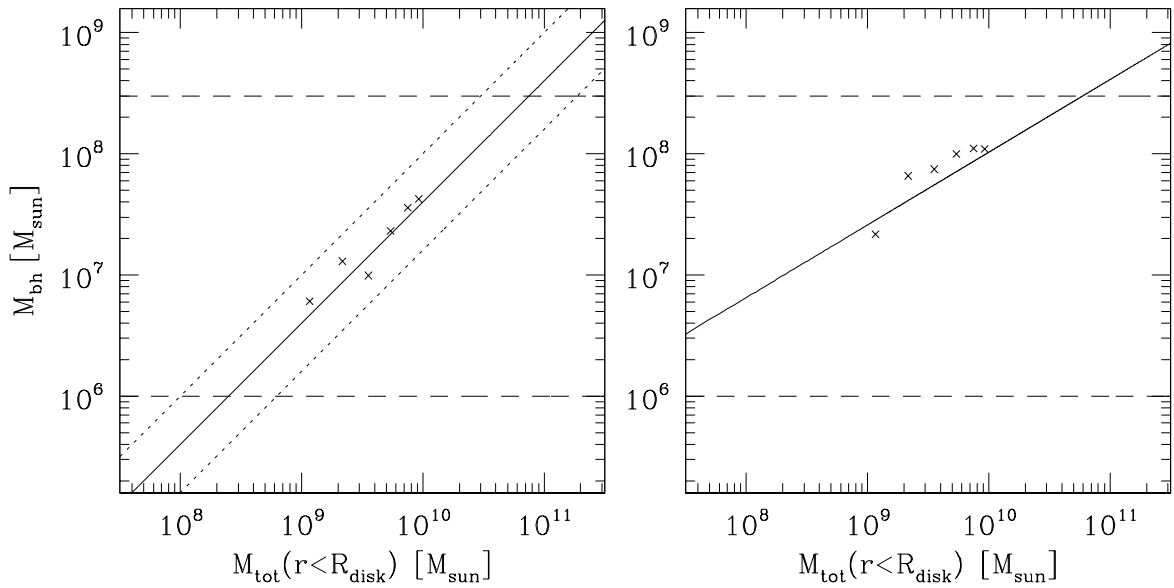}
\caption{(a) Final MBH mass at the time t = 7 Myr, plotted against the total dynamical mass $(\rm M_{gas} + M_{star})$ enclosed within the initial disk, for the six different runs (black crosses). The solid line corresponds to $\rm M_{BH} \propto M_{tot}$ and the dotted lines show the typical observational uncertainty of the  `$\rm M_{BH}$ - $\rm M_{virial}$'   relation. (b) Same as (a) but for the final MBH mass at the end of all simulations. The solid line corresponds to $\rm M_{BH} \propto M_{tot}^{0.7}$.  In both figures, the dashed lines shows the possible minimum $\rm (M_{bh}(t=0) = 10^{6} M_{\sun})$ and maximum $\rm (M_{gas}(t=0) = 3 \times 10^{8} M_{\sun})$ final MBH mass.}
\label{f4}
\end{figure}

Finally,  we plot the final black hole mass for the six different runs at  t = 7 Myr in Fig. $\ref{f4}$a and at the end of all simulations in  Fig. $\ref{f4}$b. In each run,  the host spheroid has a different  mass  and therefore a different total dynamical  mass $(\rm M_{star} + M_{gas})$ enclosed within the initial disk, as plotted in Figs $\ref{f4}$a and b. Fig. $\ref{f4}$a shows a clear correlation between the MBH mass and the host spheroid mass. The solid line is $\rm M_{bh} \propto M_{tot}$ and the scatter is considerably less than the `$\rm M_{BH}$ - $\rm M_{virial}$'  relation. Fig. $\ref{f4}$b shows that the correlation tends to get tilted towards $\rm M_{bh} \propto M_{tot}^{0.7}$ (solid line in  Fig. $\ref{f4}$b). As mentioned before, our simulations suffered of an unrealistic  steep decay  in the star formation rate after t = 7Myr. This effect is more notorious at spheroids with lower  masses because they consume gas less efficiently than bigger spheroids (their disks are less turbulent and therefore with less inhomogeneities), having more leftover gas to be mainly consumed by the MBH after t = 7Myr and thus tilting the relation. Therefore we think that the more reliable result is in the one up to t = 7Myr (Fig.$\ref{f4}$a) and is the result that we will continue analyzing  in this paper.  

We think that the little scatter observed in Fig. $\ref{f4}$a  is expected taking into account  that our simulations are still idealized, with only one parameter varied  and that only tries to reveal the physical process (fueling)   responsible of the `$\rm M_{BH}$ - $\rm M_{virial}$'  relation. Many    other processes  that are not taken into account in this work, such the number of major mergers (growth episodes) that happened in the history of a given spheroid, the feedback from star formation and possibly from an AGN,  the fraction of the mass fed that ends up inside the event horizon of the MBH, within other processes    will contribute to increase the scatter in the relation. 
 
\section{Generalized  Kennicutt-Schmidt Law and the Origin of the Relation}

 Using simple analytical models, E06  shows that the average mass  fueling onto the inner accretion disk $\dot{<M>}$ is proportional to $\sigma^{3}$ (eq 3 in E06). This result is valid for the different possible types of massive nuclear disk, and also different mechanisms for angular momentum transport, being  the zero-point the only way to discriminate between them (see \S 5 in the same paper). E06 also   couple this result with the expected gas lifetime given by the  Kennicutt-Schmidt Law (Kennicutt (1998); $\rm  t_{gas} \approx t_{SF} \, =  \,  \Sigma_{gas}/\dot{\Sigma}_{SF} \, \propto \, t_{orb} = R_{d}/\sigma \,\,$), naturally leading to the `$\rm M_{BH}$ - $\rm M_{virial}$' relation:
\begin{equation}
\rm M_{BH} = \dot{M} \, t_{gas}  \propto R_{d}\sigma^{2} \propto  M_{virial}.
\end{equation}

\begin{figure}
\plotone{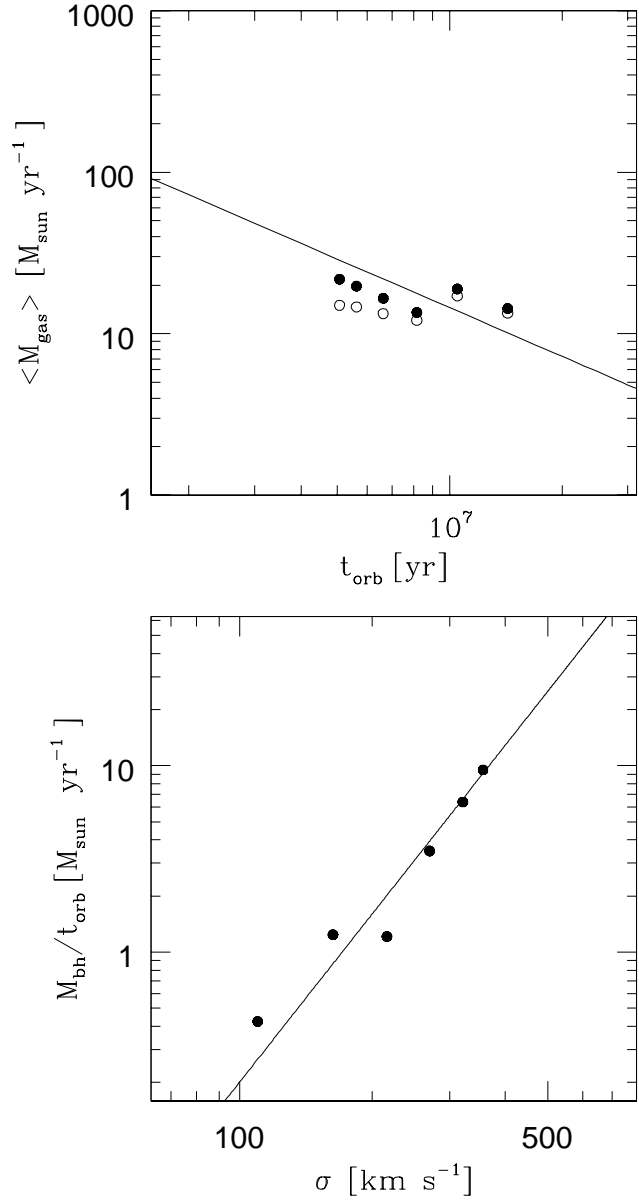}
\caption{(a) Average gas depletion rate  $\rm < \dot{M}_{gas}>$ as a function of the orbital time, for the six different runs. The open circles are the gas depletion rate due to only star formation, the filled circles are the gas depletion rate due to both star formation and accretion onto the MBH. The solid  line corresponds to  $\rm < \dot{M}_{gas}> \, \propto \, t_{orb}^{-1}$ (b) Total mass accreted by the MBH per orbital time (filled circles) plotted against the stellar velocity dispertion $\rm \sigma$, for the different runs.  The solid  line corresponds to  $\rm M_{BH}/t_{orb} \, \propto \, \sigma^{3}$ }
\label{f5}
\end{figure}

In this section, we check if both conditions are satisfied in our simulations. We start  by studing if the star formation rate in our model satisfies the  Kennicutt-Schmidt Law: $\rm \dot{\Sigma}_{SF}/\Sigma_{gas} = \dot{M}_{SF}/M_{gas} \propto  t_{orb}^{-1}$. Figure \ref{f5}a shows the average gas depletion  rate $\rm <\dot{M}_{gas}>$ againts the orbital time ($\rm = 2\pi R_{d}/\sigma$) for the six different runs. Since the total gas mass  $\rm M_{gas}$ is the same for all the runs, the Kennicutt-Schmidt Law predicts that  $\rm <\dot{M}_{gas}>$ must be proportional to $\rm 1/t_{orb}$ like the solid line in Fig. \ref{f5}a. The open circles are the gas depletion rate including only the gas depleted by star formation as in the  Kennicutt-Schmidt Law. The filled circles   are again the gas depletion rate for each simulation, but now including the    gas depleted by star formation  and by  accretion onto the central black hole. The figure clearly shows that is the total gas depletion rate the one that correlates better with the inverse of the orbital time  $\rm t_{orb}$, and not the  gas depletion rate due to   star formation only.


Kennicutt (1998) estimates the star formation rate by their far infrared luminosity ($\rm L_{FIR}$). It is well known that $\rm L_{FIR}$ has a relevant AGN contribution (approximately 1/3), therefore what  Kennicutt (1998) identifies as star formation rate  is in fact closer to the total  gas depletion rate as in our simulations. E06 also neglect the bh's mass accretion rate in the gas depletion timescale, making $\rm  t_{gas} \approx t_{SF} \propto t_{orb}$. Our simulations shows that such aproximation is no longer needed, because is the total (accretion plus star formation) gas depletion timescale itself the one that scales with the orbital time. 

The  gas depletion rates found in our simulations are higher than Kennicutt (1998) ones by a factor of $\sim$ 5. This is an expected result since any feedback or recycling of gas has been included in our simulations. Only some feedback is implicitly represented  by the artifitial cut-off of the cooling funtion at T = 10,000 K. But this only corresponds con an effective sound speed $\rm c_{s} \approx 10 km \, s^{-1}$ that is well below compared to the degree of turbulence found in starburst galaxies. 

The exact slope of this kinematical formulation of the Kennicutt-Schmidt Law,  $\rm \dot{\Sigma}_{SF}  \propto (\Sigma_{gas} / t_{orb})^{\alpha}$ with $\rm \alpha = 1$ being the original Kennicutt (1998) result, has been topic of debate in the recent years. Boisser et al. (2003) reported a steeper slope ($\rm \alpha = 1.5$) from the observation of 16 normal disk galaxies and this steeper  slope  has been recently confirmed numerically  (Li, Mac Low \& Klessen 2006). The origin of this discrepancy has been attribute to a difference between normal galaxies and starburst galaxies. Moreover, a more carefull examination of Fig 7 in  Kennicutt (1998), also shows that the normal galaxies considered alone seem to have a steeper slope that  starburst ones (Li, Mac Low \& Klessen 2006). Since our simulations are only in the starburst regime, our work is in agreement with   a less steep  slope of $\rm \alpha = 1$ for starburst galaxies.

The second  condition needed to be satisfied by our simulations is that  $\rm \dot{<M>}$ is proportional to $\sigma^{3}$, something that was already tested in E06 for an adiabatic disk. We estimated  $\rm \dot{<M>}$ by the mass accretion rate onto the sink particle per orbital time $\rm M_{BH}/t_{orb}$. Fig  \ref{f5}b shows  $\rm M_{BH}/t_{orb}$ as a function of $\sigma$, where the filled circles are  the mass accretion rate onto the sink particle per orbital time for the different runs and the solid line is  $\rm   \dot{<M>} \propto \sigma^{3}$. Therefore the   condition   $\rm   \dot{<M>} \propto \sigma^{3}$ is also satisfied 
 by our simulations.

E06 numerically found that  $\rm \dot{<M>}$ is proportional to $\sigma^{3}$ for a `gravito-driven' turbulent disk, where both turbulent viscosity and gravitational torques are  the  responsible for the transport of mass. These simulations can be  more characterized by  what is called a `disk of clouds', where only gravitational torques are  the  responsible for the transport of mass and suggesting that this  is a fairly general result. In fact it is endeed a general result, because for any given disk rotating at a  speed $\rm v_{rot}$, simple dimensional analysis tells us that the mass accretion rate $\rm \dot{M}$ will scales as   $\rm v_{rot}^{3}/G$. This is the dimensional dependence of $\rm \dot{M}$ independent of the proccess responsible for  angular momentum loss and  mass transport, that will only change the dimensionless dependence. Taking into account that in ULIRGs  $\rm v_{rot} \approx \sigma$(Downes \& Solomon 1998; see \S 2.1 in E06) within aproximatelly 10\%, we arise to the expected result of    $\rm \dot{M}$ being proportional to $\sigma^{3}$.

\section{Fundamental Link Between Massive Black Holes and Host Galaxies}

In the recent years, two important correlations has been found between  massive black holes and the host spheroid (bulge/elliptical galaxies) properties. The first one is that  the masses of black holes correlates with  the mass of the host ($\rm M_{bh}-M_{bulge}$; Magorrian et al. 1998), and in the second one their masses correlates with average random velocities of the stars in their host ($\rm M_{bh}-\sigma$; Ferrarrese \& Merritt 2000; Gebhardt et al. 2000). Since both relations are connected by the Faber-Jackson relation   $\rm R_{e} \propto \sigma^{2-3}$, only one relation is the real connection between MBHs and galaxies. In other words, the other relation is then a mere byproduct of the real one and the  Faber-Jackson relation.

Initially, the $\rm M_{bh}-\sigma$ appears with considerably less scatter suggesting that either  is the more fundamental relation or  the estimation for $\rm M_{bulge}$ used in  Magorrian et al. (1998), that was simply the host luminosity $\rm L_{bulge}$, was a poor  $\rm M_{bulge}$ estimator compared to  $\rm \sigma$. Subsequent work with a more accurate estimate for $\rm M_{bulge}$ (Marconi \& Hunt 2003; Haring \& Rix 2004; Lauer et al. 2006), endeed shows that the $\rm M_{bh}-M_{bulge}$ relation has comparable or less scatter than the $\rm M_{bh}-\sigma$ relation. Therefore is still unclear which relation is the fundamental link between MBH and host spheriod and which is a byproduct of the other one.

\begin{figure}
\plotone{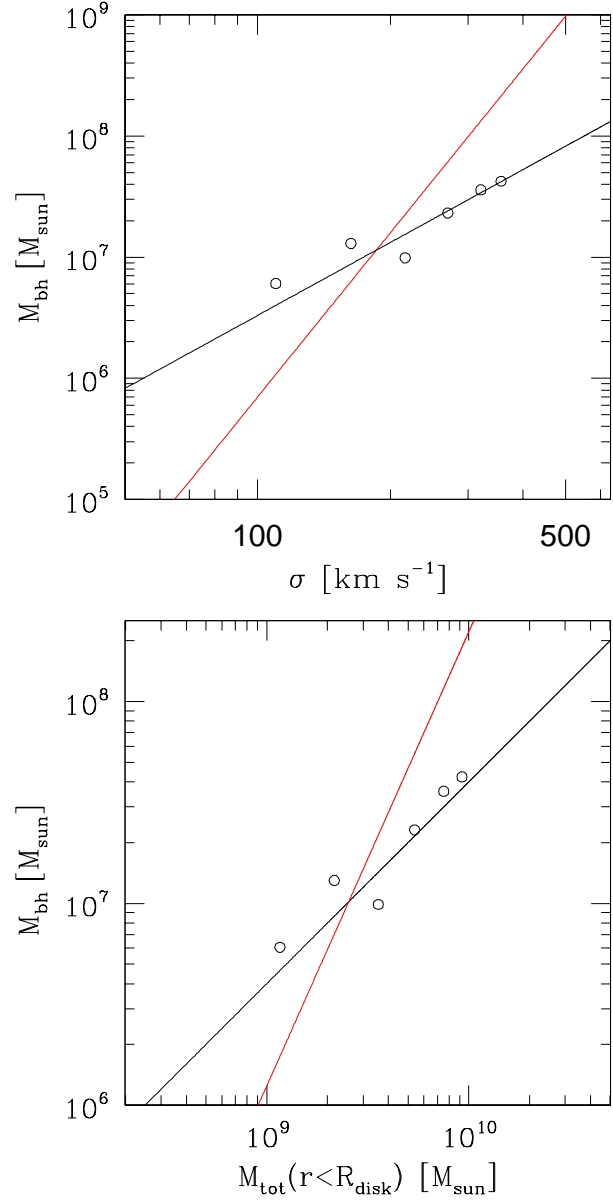}
\caption{(a) The figure shows the final MBH mass  plotted  against the stellar velocity dispersion $\rm \sigma$  for the six different runs (open circles). The black curve corresponds to $\rm M_{BH} \, \propto \, \sigma^{2}$ and the red curve corresponds to $\rm M_{BH} \, \propto \, \sigma^{4.5}$. (b)  Final MBH mass  plotted now against the total dynamical mass $(\rm M_{gas} + M_{star})$ enclosed within the initial disk (open circles). The black curve corresponds to $\rm M_{BH} \, \propto \, M_{tot}$ and the red curve corresponds to $\rm M_{BH} \, \propto \, M_{tot}^{2.25}$.}
\label{f6}
\end{figure}

Our simultion setup  has the advantage of a fix initial disk radius $\rm R_{disk}$ in all runs that decouples both relations. Since for a spherically symetric system, only the enclosed stellar mass by the intial disk has dynamical influence onto the disk,  $\rm R_{disk}$  defines the  mass dynamically relevant ($\rm M_{dyn} \approx  \sigma^{2}R_{disk}/G$) instead of  $\rm R_{e}$ (that is $ \propto \sigma^{2-3}$; Faber-Jackson relation). Therefore, in  our configuration  we have instead one set of  two  consistent solutions (`$\rm M_{BH}$ - $\sigma$' \& `$\rm M_{BH}$ - $\rm M_{virial}$'  relations), we have two possible sets of  consistent solutions:

\begin{equation}
\rm M_{bh} \propto M_{virial} \Rightarrow  M_{bh} \propto \sigma^{2}
\label{sol1}
\end{equation}

\begin{equation}
\rm M_{bh} \propto \sigma^{4.5} \Rightarrow  M_{bh} \propto M_{virial}^{2.25} 
\label{sol2}
\end{equation}
In Fig \ref{f6}a we plot the black holes masses for the six different runs, which corresponds to different  $\sigma$. Fig \ref{f6}b  plots again the black holes masses for the different runs, but now plotted against the total enclosed dynamical mass. The black lines are the predictions for the first set of possible solutions (Eqs. \ref{sol1}) and the red lines are the predictions for the second ones (Eqs. \ref{sol2}). As  expected from the results showed earlier in this paper, this figure clearly shows that the runs lies in the first set  of possible solutions (Eqs. \ref{sol1}). This means  that in a fueling-controlled growth scenario, the  `$\rm M_{BH}$ - $\rm M_{virial}$'  relation is the fundamental link between MBH and host spheroids, and the `$\rm M_{BH}$ - $\sigma$' relation is only a byproduct of this  relation.

The  relevance  of  the `$\rm M_{BH}$ - $\rm M_{virial}$'  relation was already predicted in  E06 and is in agreement with recent extension of the correlations to lower masses. It has been found by Ferrarrese et al. (2006) and Wehner \& Harris (2006) that compact stellar nuclei correlates with the masses of the host galaxies, falling along the same scaling relation that MBHs obeyed with their, predominantly more massive, host galaxies. Contrarily, when they plot their masses against their host velocity dispersion $\rm \sigma$, the nuclei and MBHs obey statiscally different scaling relation suggesting that is not the fundamental byproduct of the proccess of galaxy formation. In our simulations, the central sink particle could be either a MBH or  a  compact stellar nuclei depending on how the mass evolves on this un-resolved scales. But its total mass must obey the same scaling relation that is established by the larger scales gravitational dynamics of fueling and star formation at the epoch of galaxy formation.

\section{Discussion}

Since their discovery, to many  analytical explanations for the `$\rm M_{BH}$ - $\rm M_{virial}$'  and `$\rm M_{BH}$ - $\sigma$' relations has been proposed. In these theories are represented  almost all kind of distinct physical processes, giving the impresion that the relations are not a strong constrain on the coevolution of MHBs and host galaxies. However, many of them are simply not physically plausible  and others use key hypothesis that must    be proven correct first. Therefore in order to validate theses theories, they need to be  numerically tested. This can only be done with 
 numerical simulations that are expecially designed to test their key hypothesis or physical processes and that has enough  resolution  to do it. This work  is our  first  attempt to validate numerically  that the origin of  the `$\rm M_{BH}$ - $\rm M_{virial}$'  is in the mass  fueling onto the inner accretion disk.

Since fueling of mass onto the inner accretion disk is already an unavoidable step in the growth of MBHs, the theory presented here  arise as a natural explanation for the  `$\rm M_{BH}$ - $\rm M_{virial}$' relation. Also, has the advantage that this theory don't  needed any extra ad-hoc hyphotesis such as huge outflows, fine tunning on the gas mass to stellar mass ratio, etc in order to explain the observed correlations. Contrarily,  fits quite well in the standard theory of galaxy formation, where spheriods are formed in rapid star forming events triggered by galaxy mergers.  

This  work considerably differs from previous numerical work on the growth of MBHs that mainly focus on larger (galaxy mergers) scales. For example in Di Matteo et al. (2005) the gravitational forces are fully resolved only on scales larger than 1 Kpc, but below this  scales is where the most relevant  processes (ongoing starburst, black hole accretion and feedback from both) happens. Therefore they rely all this relevant physics on simple parametrized prescriptions, making their  approach not far from the semi-analytical one. Moreover, they attribute  the origin of the correlations to   AGN feedback (parametrized too), but they don't  give enough evidence for that. Li et al (2006) does  a  better job in terms of resolution, with their $\rm \sim 40$ pc gravitational and sink particle resolution they can resolve gravitational collapse and fragmentation on larger scales. However, the resolution  is still not enough to follow the fueling onto the accretion disk and they don't analyze   enough on the origin of the correlations.

In this work we don't follow the full galaxy merger and instead we assume a massive nuclear disk at the center of the remnant, but such disk  is a well known by-product of a gas-rich galaxy merger (Barnes \& Hernquist 1996; Di Matteo et al. 2005; Mayer et al. 2007). We take a more experimental or exploratory approach where give  some hyphotesis, $\rm \dot{M} \propto \sigma^{3}$ and  $\rm t_{gas} \propto t_{orb}$ ,  to be tested and designed  a set of simulations that  succesfully test  them under its  limitations. Our simulations have of course idealizations, such as not including feedback processes from star formation and AGN. We think that they do play a role in the physical state of the ISM, especially feedback from star formation, but we don't think that they are responsible for origin of the correlations since they a natural byproduct of mechanisms that we know must be present (BH's fueling and star formation). Finally since these mechanisms are already present, other alternative mechanisms must erase the already existing correlations first and make them  again, something that is  possible but very unlikely.


One of the major uncertainties remained is what is the fate of the gas within the accretion disk, which is another big problem by itself. Unfortunately, is still unclear how much of the mass fed onto the inner accretion disk ends up inside the event horizon of the MBH. It is also unclear the dynamical effect that will have the non accreted material in the incoming accretion flow. However besides these uncertainties we  think that they should not  change considerably the overall picture, like for example destroying  the correlations with the galactic properties settled in the fueling process, since these correlation are endeed observed. Using simple models, E06 estimate that  only around 15\% of the mass fueled is needed to be accreted in order to reproduce the observed `$\rm M_{BH}$ - $\rm M_{virial}$' relation.


\section{Summary}

In our previous paper (E06) we studied the relation between nuclear massive black holes and their host spheroid gravitational potential. Using simple  models, we analyzed how gas is expected to be transported in the nuclear regions of galaxies. When we  coupled it with the expected gas lifetime given by the  Kennicutt-Schmidt Law, naturally lead us to the `$\rm M_{BH}$ - $\rm M_{virial}$'  and `$\rm M_{BH}$ - $\sigma$' relations. 

In the present paper, we have extended this work based on simple  model by studying    considerably  more realistic adaptive mesh refinement simulations using the  ENZO code. We use this powerful tool to study how gas is  transported in the nuclear (central kpc) regions of galaxies down to  the inner accretion disk (sub-pc scale). We also study  star formation in this massive nuclear disk that  is an ubiquitous feature  in  proto-spheroids (ULIRGs, SCUBA Galaxies).

We found that  these sub-pc resolution simulation of gas  fueling that is mainly depleted by star formation naturally satisfy  the `$\rm M_{BH}$ - $\rm M_{virial}$'   relation, with a  scatter much less than in the observed correlation. 


We also found that instead of being the  Kennicutt-Schmidt Law satisfied, a generalized version of it is satisfied. In this generalized law is the total gas consumption rate $\rm \dot{M}_{gas} \, = \, \dot{M}_{BH} \, + \,  \dot{M}_{SF}$,  the one that scales as $\rm M_{gas}/t_{orbital}$. We argued that what  Kennicutt (1998) identifies as star formation rate  is in fact closer to the total  gas depletion rate as in our simulations.

Finally, we take advantage of the particular desing of our simulations to compare the `$\rm M_{BH}$ - $\sigma$' and `$\rm M_{BH}$ - $\rm M_{virial}$'  relations. We  found that the `$\rm M_{BH}$ - $\sigma$' relation is a byproduct of the `$\rm M_{BH}$ - $\rm M_{virial}$'  relation, being therefore the latter  the most fundamental one.

\bigskip

I thank Tom Abel for his introduction to ENZO and for stimulating discussions. I also thank Roger Blandford for very valuable comments. I performed these calculations on 16 processors of a SGI Altix 3700 Bx2 at KIPAC at Stanford University. 


\newpage

\end{document}